\documentstyle[epsfig,sprocl]{article}

\def\be{\begin{equation}}
\def\ee{\end{equation}}
\def\bea{\begin{eqnarray}}
\def\eea{\end{eqnarray}}
\newcommand{\ba}{\begin{array}}
\newcommand{\ea}{\end{array}}
\def\bra{\langle}
\def\ket{\rangle}
\def\a{\alpha}
\def\b{\beta}

\def\to{\rightarrow}

\begin{document}

\begin{flushright}
KAIST-TH 98/19 \\
SNUTP 98-111
\end{flushright}

\bigskip

\title{Contribution of $b\rightarrow s gg$ through the QCD anomaly 
\\ in exclusive decays $B \rightarrow (\eta^{'},\eta) (K, K^{*})$ }

\author{P. Ko \footnote{Invited talk at the Fourth International 
Workshop on Particle Physics Phenomenology at Kaohsiung, Taiwan, 
June 18-21, 1998}}

\address{ Department of Physics
\\ Korea Advance Institute of Science and 
Technology \\ Taejon 305-701, KOREA \\E-mail: pko@charm.kaist.ac.kr} 

%


\maketitle\abstracts{
The charm content in $\eta^{'}$ is estimated using the QCD $U(1)$ axial 
anomaly applied to $b\rightarrow s g g \rightarrow \eta^{'} s$. 
We find $f_{\eta^{'}}^{(c)} \approx - 3.1$ MeV for $m_c = 1.3$ GeV. 
Our estimate agrees with other independent methods by Feldmann and Kroll, 
and Araki, Musakhanov and Toki. The resulting branching ratios for $B 
\rightarrow \eta^{'} K$ in the generalized factorization model are marginally 
consistent with the CLEO data, although they are lower than the data by 
$\sim 2$. 
}

\section{Introduction}

In the year of 1997, the CLEO collaboration reported measurements in a 
number of exclusive two-body non-leptonic decays of the type $B\rightarrow
h_1 h_2$, where $h_1$ and $h_2$ are light mesons and the inclusive
decay $B^{\pm} \rightarrow \eta^{\prime}X_s$. \cite{CLEO1}$^{-}$
\cite{CLEO2} In particular,  large branching ratios into the final
states including $\eta^{\prime}$ are reported : \cite{CLEO1}
\begin{eqnarray}
\label{etapkpm}
BR ( B^{\pm} \rightarrow \eta^{\prime} + X_s ) &  = & (6.2 \pm 1.6 \pm
1.3 ) \times 10^{-4} 
\\
& & ({\rm for} ~2.0~{\rm GeV} \le p_{\eta^{'}}  \le
2.7~ {\rm GeV}),
\nonumber  
\\
BR ( B^\pm \rightarrow \eta^{\prime} + K^\pm) & = & (7.1_{- 2.1}^{+ 2.5}
\pm  1.0) \times 10^{-5},
\\
BR (B^0 \to \eta^{\prime} K^0 ) & = & (5.3_{-2.2}^{+2.8} \pm 1.2) \times 
10^{-5}~. 
\end{eqnarray}
They did not observe any decay involving the $(\eta K)$ or 
$(\eta,\eta^\prime)K^*$ modes.
These measurements have stimulated a lot of theoretical activity, 
\cite{AG97}$^{-}$\cite{Kim97} both in the inclusive and exclusive decays 
involving $\eta^{'}$ mesons.

Such unexpectedly large branching ratios of $B$ decays into the final 
states with an $\eta^{'}$ meson led to  an idea that the charm content 
in $\eta^{'}$ meson might be very large.  
The amplitude for  $b \to s (\bar{c} c) \to s (\eta^{\prime},\eta)$ 
can be parametrized as $\langle \eta^{\prime} 
|\overline{c}\gamma_{\mu} \gamma_5 c|0\rangle =
-if_{\eta^{\prime}}^{c} q_{\mu}$, and similarly for the $\eta$ case. 
The relevant quantity $f_{\eta^{(\prime)}}^{c}$ is often referred to 
as the charm content of the  $\eta^{(\prime)}$. 
This quantity is {\it a priori} not known, but it was claimed that the
data (1)-(3) can be accomodated if $f_{\eta^{\prime}}^{c}$ is as large as 
$(50-180)$ MeV. \cite{HZ97,SZ97}   
However, this is uncomfortably large compared to the typical
meson decay constants, {\it e.g.}, $f_{\pi} = 132$ MeV. 
Later it turned out that they could be determined in 
a number of different ways, also including the $B$-decays combined with 
the QCD $U_{\rm A}(1)$ anomaly \cite{ACGK} being discussed here. 

In this talk, I'd like to present another method for computing the 
contribution of the amplitudes 
$b\rightarrow s (gg) \rightarrow s(\eta^{\prime},\eta)$. 
This method is based on calculating the amplitude for the chromomagnetic 
penguin process
$b\rightarrow s gg$, followed by the transitions $gg\rightarrow
(\eta^{\prime}, \eta)$ which are calculated using the
QCD anomaly, determining both the sign and magnitude of
these contributions. 
The branching ratios for $B^\pm \to (\eta^\prime,\eta)(K^\pm,K^{*\pm})$
and $B^0 \to (\eta^\prime,\eta)(K^0,K^{*0})$ 
based on the QCD-anomaly method are calculated in this letter and 
compared with the present CLEO measurements and with the ones in
Ref.~[5]. We find that the theoretical branching ratios for $B^\pm
 \to \eta^\prime  K^\pm$ and
$B^0 \to \eta^\prime K^0$ are almost equal and both are in the range
 $(2 -4)  \times 10^{-5}$, in agreement with the estimates in
Ref.~[5].  

\section{Estimate of ${\mathbf b \to (\eta,\eta^\prime)s}$ via QCD
anomaly} 

Let us consider 
the charm-anticharm pair into two gluons, followed by the transition
$gg \rightarrow \eta^{(')}$ (see Fig.~1). The first part of
this two-step process, i.e. $b \rightarrow s
(\bar{c} c \rightarrow g(k_1) g(k_2))$
which amounts to calculating the charm-quark-loop from which two
gluons are emitted, has been worked out by Simma and Wyler
\cite{simma} in the context
of a calculation in the full theory. Their result is readily
translated to our effective theory approach and can be
compactly written as a new (induced) effective Hamiltonian
$H_{\mathrm{eff}}^{gg}$, 
\begin{eqnarray}
  \label{newop}
H_{\mathrm{eff}}^{gg} & = & - \frac{\alpha_s}{2\pi} \left(
C_2^{\mathrm{eff}} + \frac{C_1^{\mathrm{eff}}}{N_c} \right)
\frac{G_F}{\sqrt{2}} V_{cb} V_{cs}^* \Delta i_5
\bigl(\frac{q^2}{m_c^2}\bigr)
\frac{1}{k_1 \cdot k_2} 
\nonumber \\
& \times & G^{\alpha \beta}_a (D_{\beta}
\tilde{G}_{\alpha \mu} )_a \, \overline{s}\gamma^{\mu} (1-\gamma_5) b
\quad , 
\end{eqnarray}
with $\tilde{G}_{\mu \nu}=\frac{1}{2} \epsilon_{\mu \nu \a \b} G^{\a
\b}$ $(\epsilon_{0123}=+1)$. The $C_{1,2}^{\rm eff}$ are modified Wilson 
coefficients for the current-current operators as described in 
Ref.~[5]. 
In this formula, which holds for on-shell gluons
($q^2 = (k_1 + k_2)^2 = 2k_1 \cdot k_2$),
the sum over color indices is understood. The function $\Delta
i_5 (q^2/m_c^2)$ is defined as 
\begin{equation}
  \label{deltai5}
  \Delta i_5 (z) = -1 +\frac{1}{z} \Bigl[ \pi -2 \arctan (\frac{4}{z}
  -1)^{1/2} \Bigr]^2, \ \mbox{for $0<z<4$} \quad .
\end{equation}
Note that the $b \to s gg$ calculation brings in an explicit factor of
$\alpha_s$.  However, this explicit $\alpha_s$ factor gets 
absorbed into the matrix element of the operator resulting from  the anomaly. 
So, to the order that we are working, we use the coefficients 
$C_1^{\mathrm{eff}}$ and $C_2^{\mathrm{eff}}$ in Eq.~(\ref{newop}).
The $u\bar{u}$ contribution in
Fig. 1 is suppressed due to the unfavourable CKM factors. The 
$t\bar{t}$ contribution is included in the effective Hamiltonian via 
the $bsgg$ piece present in the operator $O_g$. However, in the
factorization framework, the $bsgg$ term in $O_g$ does not contribute
to the decays discussed. So, the $c\bar{c}$ contribution in Fig. 1 is
the only one that survives.

Working out the hadronic matrix element of Eq.~(\ref{newop})
using factorization, we now need to evaluate the matrix
elements : 
\be
\langle \eta^{(\prime)} |  
G_a^{\alpha \beta} (D_{\beta} \tilde{G}_{\alpha \mu, a}) |0\rangle, 
\ee 
which can be written as
\begin{equation}
  \label{parder}
 G^{\alpha \beta}_a (D_{\beta} \tilde{G}_{\alpha\mu} )_a =
 \partial_{\beta} (G^{\alpha \beta}_a \tilde{G}_{\alpha \mu,a}) -
 (D_{\beta} G^{\alpha \beta})_a \tilde{G}_{\alpha \mu,a} .
\end{equation}
Now we can discard the second term since it is suppressed by an additional 
power of $g_s$ which follows on  using the equation of motion, and
furthermore, the first 
term is enhanced by $N_c$ in the large $N_c$ limit. The matrix elements
of $\partial_{\beta} (G^{\alpha \beta}_a \tilde{G}_{\alpha \mu,a})$ are
related to those of $G\tilde{G}$ :
\be
\partial_\beta \bra \eta^{(')}|G^{\alpha \beta}_a \tilde{G}_{\alpha
\mu,a} | 0 \ket = \frac{i q_\mu}{4}
\bra \eta^{(')}|G^{\alpha \beta}_a \tilde{G}_{\alpha \beta,a}|0 \ket
\quad .
\ee
The conversion of the gluons into $\eta$
and $\eta^{\prime}$ is described by an amplitude which is fixed by the
$SU(3)$ symmetry and the axial $U(1)$ current triangle anomaly. The
matrix elements for $G\tilde{G}$ can be written as \cite{voloshin80}
\be
\label{anomaly}
\langle \eta^{(')} | \frac{\alpha_s}{4\pi} G^{\alpha \beta}_a
\tilde{G}_{\alpha \beta,a} |0 \rangle = m_{\eta^{(')}}^2
f_{\eta^{(')}}^u \quad .
\ee
In Eq.~(\ref{anomaly}) the decay constants $f_{\eta^\prime}^u$ and
$f_{\eta}^u$ read
\begin{equation}
\label{koppetas}
f_{\eta}^u = \frac{f_8}{\sqrt{6}} \, \cos \theta_8 -
               \frac{f_0}{\sqrt{3}} \, \sin \theta_0  \quad , \quad
f_{\eta^\prime}^u = \frac{f_8}{\sqrt{6}} \, \sin \theta_8 +
               \frac{f_0}{\sqrt{3}} \, \cos \theta_0  \quad , \quad
\end{equation}
where the coupling constants $f_8$, $f_0$ and the mixing angles 
$\theta_8$ and $\theta_0$ have been introduced earlier. 
We follow here the  two-angle $(\eta,\eta^\prime)$ mixing formalism of 
Ref.~[21], where the mass eigenstates $|\eta\rangle$ and 
$|\eta^\prime \rangle$ have the following decompositions: 
\begin{eqnarray}
\label{mix1}
|\eta \rangle = \cos \theta_8 |\eta_8 \rangle - \sin \theta_0 |\eta_0
\rangle , 
\nonumber\\ 
|\eta^\prime \rangle = \sin \theta_8 |\eta_8 \rangle + \cos \theta_0
|\eta_0  
\rangle . \end{eqnarray}
Collecting the individual steps,
the matrix elements in Eq.~(6) can be written as
\be
\label{opnewa}
\langle
\eta^{(\prime)}(q) |  \frac{\alpha_s}{4\pi}
G_a^{\alpha \beta} (D_{\beta} \tilde{G}_{\alpha \mu
a}) |0\rangle = i q_\mu \frac{m_{\eta^{(')}}^2}{4}
f_{\eta^{(')}}^u  \quad .
\ee
One would have naively expected that the gluonic matrix elements are
small since they contain an extra factor of $\alpha_s$. However, as
shown by  Eq.~(\ref{opnewa}), this is obviously not the case and the
gluon operator with $\alpha_s$ as a whole is responsible for the
invariant mass of the  $\eta^{(\prime)}$ mesons. Also, the combination
entering in Eq.~(\ref{opnewa}) involving the product of
$\alpha_s$ and the gluon field operators is independent of the
renormalization scale.

Before closing this section, let me comment on the alternative method to 
estimate $f_{\eta^{'}}^c$ using the $\eta_c - \eta^{'}$ mixing 
(Fig.~\ref{parton} (a)) :
\begin{equation}
J/\psi \rightarrow \eta_c^* + \gamma
\end{equation}
For example, in Ref.~[5], these quantities were determined
from the decays $J/\psi  \to (\eta,\eta^{\prime},\eta_c) \gamma$,
extending the usual $(\eta,\eta^{\prime})$-mixing formalism
\cite{GK87} to the $(\eta_c,\eta^{\prime},\eta)$ system. Using the
measured  decay widths for the decays 
$J/\psi \to (\eta,\eta^{\prime},\eta_c) \gamma$ 
and $(\eta_c,\eta^{\prime},\eta) \to \gamma \gamma$ yields
$|f_{\eta^\prime}^c| \simeq 5.8$ MeV and $|f_{\eta}^c| \simeq 2.3$ MeV.  
\cite{AG97} However, in such a scheme, the intermediate $\eta_c$ is far 
off-shell, and one has to know the form factor for 
$J/\psi - \eta_c - \gamma$ vertex. Moreover, this
picture is not consistent with electromagnetic gauge invariance, 
since at the quark level, the above transition occurs through 
$c\bar{c} \rightarrow \gamma + g g $ followed by $g g \rightarrow \eta^{'}$ 
(Fig.~\ref{parton} (b)).  Note that there is a diagram in which the photon is 
emitted in the middle of two gluons.  Such diagram cannot be represented 
by the simple $\eta_c - \eta$ mixing picture.  
One cannot simply neglect this diagram either, since it would violate 
electromagnetic gauge invariance. Therefore, the usual picture of 
$\eta_c - \eta^{'}$ mixing for $J/\psi \rightarrow \eta^{'} \gamma$ may 
not be a good one, although it is widely used in the literature. 

\section{Estimate of the decay rate for  ${\mathbf B \to
(\eta^\prime,\eta)(K,K^{*})}$ }

In order to compute the complete amplitude for the exclusive decays, 
one has to combine the contribution from the decay 
$b \to s (c\bar{c}) \to s(gg) \to s \eta^{(')}$ discussed in 
the previous section with all the others arising from the
four-quark and chromomagnetic operators, as detailed in Ref.~[5]. 
The resulting amplitudes in the factorization approximation are listed
in Ref.~[18]. By comparing two expressions in Refs.~[5] and 
[18], we can make the following identification :
\be
  \label{corres}
  - \Delta i_5 (m_{\eta^{\prime}}^2/m_c^2) f_{\eta^{\prime}}^u
  \rightarrow
    f_{\eta^{\prime}}^{c}, ~~~~~~~
  - \Delta i_5 (m_{\eta}^2/m_c^2) f_{\eta}^u \rightarrow
    f_{\eta}^{c}. 
\ee
Therefore, we have a simple relation between the decay constants
$f_{\eta^{\prime}}^{c}$, $f_{\eta}^{c}$, introduced in the intrinsic
charm content method, and the form factor
$\Delta i_5$ entering via the operator in Eq.~(\ref{newop}).
The idea of intrinsic charm quark
content of $\eta^\prime$ and $\eta$ and the contribution of the operator in
Eq.~(\ref{newop}) are  related since this operator
comes from the charm quark loop.
 Using the best-fit values of the
$(\eta,\eta^\prime)$-mixing parameters from Ref.~[23], yielding
$\theta_8 =-22.2^\circ,
\theta_0 =-9.1^\circ, f_8=168 ~\mbox{MeV}, f_0 =157 ~\mbox{MeV}$,
which in turn yields
$f_{\eta^\prime}^u =63.6$ MeV and $f_\eta^u =77.8$ MeV,
the relations in Eq.~(\ref{corres}) give $f_{\eta^{\prime}}^{c} \sim 
-3.1$ MeV ($-2.3$ MeV) and $f_{\eta}^{c} \sim -1.2$ MeV ($-0.9$ MeV), 
with $m_c$ having the value $ 1.3$ GeV ($1.5$ GeV). 
In our approach the relative signs of the contributions from
$O_1^c$ and $O_2^c$ to the other contributions are determined; we obtain
the negative-$f_{\eta^{\prime}}^{c}$ (and $f_{\eta}^{c}$) 
solution of the two possible ones. 
Our estimate of $f_{\eta^{\prime}}$ is consistent with the results obtained
by Feldmann, Kroll and Stech, \cite{FKS98} 
\be
f_{\eta^{\prime}} = - (6.3 \pm 0.6) ~{\rm MeV}, ~~~
f_{\eta} = - (2.4 \pm 0.2) ~{\rm MeV}, ~~~
\ee
which is based on a new scheme for the flavor mixing and the axial anomaly.
Also there appeared Ref.~[25] by Araki {\it et al.}, who did an 
independent calculation in the same way as Haperin and Zhitnitsky and got 
\begin{equation}
f_{\eta^{'}}^{(c)} = - ( 12.3 - 18.4 )~{\rm MeV},
\end{equation}
after correcting some mistakes in the orginal calculations in Ref.~
[7].  By now it is amusing that three independent methods for 
estimating $ f_{\eta^{'}}^{(c)}$ agree on its sign and the size within 
a factor of $2-3$. So one can say that the issue of the charm content in 
$\eta^{'}$ is  now settled with reasonable confidence.

We plot the resulting branching ratios $BR(B^{\pm} \rightarrow \eta^{\prime} 
K^{0 \pm})$ and $BR(B^{0} \rightarrow \eta^{\prime} K^{0})$ 
as functions of the parameter $\xi$ in Figs.~3 and 4 for three different 
sets of CKM elements $\rho$ and $\eta$ in the Wolfenstein parametrization 
: \cite{W83}   
\be
(\rho, \eta )  = (0.05, 0.36), ~(0.30, 0.42), ~{\rm and}~( 0, 0.22)
\ee
The branching ratios for the neutral $B$-meson decays are averages with the
corresponding charged conjugated decays in the figures. 
The numerical results for the branching ratios depend on various input
parameters, the details of which can be found in Ref.~[18].
Let us discuss the sensitivity on various parameters in brief.
The branching ratios show a mild dependence (of order $10\%$) on the CKM 
parameters. However the branching ratios depend on the $s$-quark mass very
sensitively, because the amplitude for $B\rightarrow \eta^{'} K$ contains 
terms with the $1/m_s$ factor. 
In this work, we used $\overline{m}_s ( 2.5 ~{\rm GeV} ) = 122$ MeV. 
However, if the present high values of the branching ratios for 
$B^{0(\pm)} \rightarrow \eta^{\prime} K^{0(\pm)}$ continue to persist, 
one might have to consider smaller values of $\overline{m_s}$.
We also expect  progress in calculating quark masses on the
lattice, sharpening the theoretical estimates presented here.
Dependence on $\xi$ amounts to between 20\% and 35\%
depending on the other parameters if one varies $\xi$ in the
range $0 \leq \xi \leq 0.5$. In all cases, the branching ratios are
larger for $\xi=0$. 
Taking into account the parametric dependences just discussed,
we note that the theoretical branching ratio 
$BR(B^{0(\pm)} \rightarrow \eta^{\prime} K^{0(\pm)})$ 
are uncertain by a factor 2.  For the ratio of the branching ratios
$BR(B^{\pm} \rightarrow \eta^{\prime} K^{\pm})/BR(B^{0} \rightarrow 
\eta^{\prime} K^{0})$,  which is  useful quantity since
it is practically independent of the form factors and most input   
parameters, the residual uncertainty is due to the CKM-parameter 
dependence of this ratio. 
It is estimated as about 10\%. We get (for $0 \leq \xi \leq 0.5$) 
\begin{equation}
\label{ratior1}
\frac{BR(B^{\pm} \rightarrow \eta^{\prime} K^{\pm})}{BR(B^{0}
\rightarrow \eta^{\prime} K^{0})} = 0.9 - 1.02 ~.
\end{equation}  
The present experimental value of this ratio as calculated by adding
the experimental errors in the numerator and denominator in quadrature
is  $1.34 \pm 0.85$. 
One can also study other modes involving $K^*$ instead of $K$.
Summarizing the results of Ref.~[18], 
the branching ratios for the decay modes $B^{\pm}\rightarrow 
\eta K^{\pm}$ and $B^{0} \rightarrow \eta K^{0}$ are smaller compared
to  their $\eta^\prime$-counterparts by at least an order of magnitude,
namely,  $ BR(B^{\pm}\rightarrow \eta K^\pm) =(1 - 2) \times 10^{-6}$ 
and a similar value for the neutral $B$ decay mode. On the other hand, 
the branching ratios for the decay modes $B^\pm \to \eta (K^\pm,K^{*\pm})$  
and $B^0 \to \eta (K^0,K^{*0})$ are all comparable to each other 
somewhere in the range $(1 - 3) \times 10^{-6}$.

\section{Concluding Remarks}

In this talk, we have presented an independent estimate of the charm content 
of $\eta^{(')}$ meson using the process $b \to s(c\bar{c}) \to s (gg) \to s 
(\eta^\prime,\eta)$ and QCD anomaly. We could fix its sign as well as the 
size. Our results are in good agreement with other results based on 
independent methods, and thus we can be confident by now that the charm
content in $\eta^{(')}$ meson is indeed small.  
One could further study the branching ratios in $B \to 
(\eta^\prime,\eta)(K,K^*)$ in the generalized factorization approximation.  
Our result is marginally consistent with the CLEO data, considering 
various theoretical uncertainties in this kind of game (Figs.~3 and 4). 
Also our result is drastically different from the ones which follow in other 
scenarios. Hence, ongoing and future experiments will be able to test the 
predictions of the present approach as well as of the
competing ones, such as models based on the dominance 
of the intrinsic charm contributions in $\eta^\prime$, as suggested in 
Refs.~[7,8], or models in which dominant role is attributed 
to the soft-gluon-fusion process to form an $\eta$ or $\eta^\prime$. 
\cite{Mohd97,Kim97}

\begin{figure}[p]
\centerline{
\epsfig{file=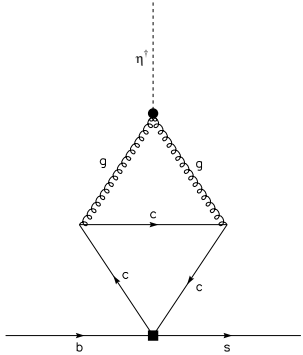,
height=2.0in,angle=0,clip=}
}
\caption[]{
Feynman diagram contributing to the processes $b \to s
(c\bar{c}) \to s (gg) \to s \eta^{(')}$ in the full and effective theory.
 The lower vertex in the diagram on the right is
calculated with the insertion of the operators $O_1^c$ and $O_2^c$ in
the effective Hamiltonian approach; the upper vertex in both the full and 
effective theory is determined by the QCD triangle anomaly. 
\label{fig1}}
\end{figure}


\begin{figure}[p]
\centerline{
\epsfig{file=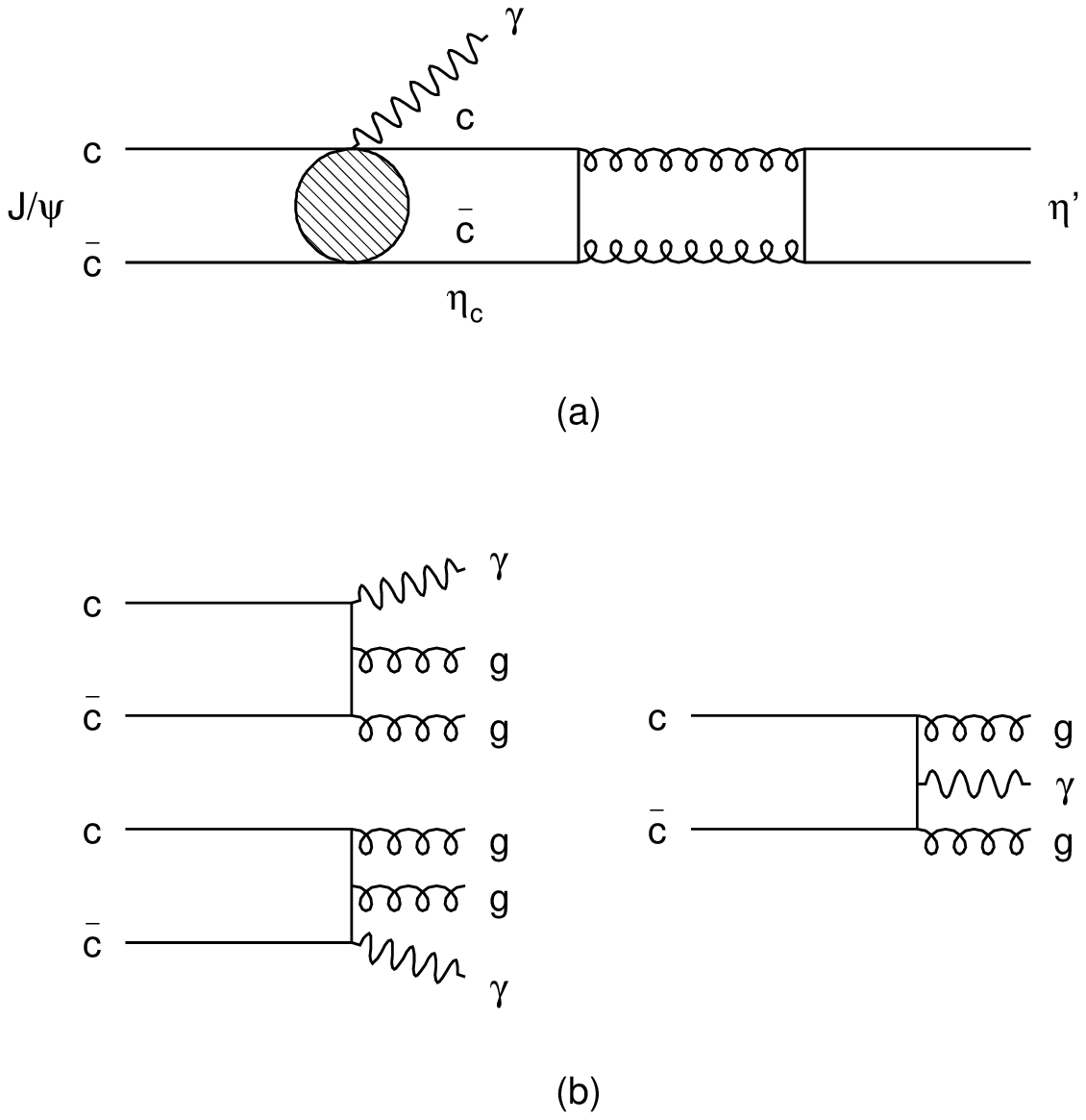,
height=3.0in,angle=0,clip=}
}
\caption[]{
Feynman diagram contributing to the processes $J/\psi \to \gamma \eta^{'}$ 
(a) in the $\eta_c - eta^{'}$ mixing picture,  and (b) in the parton 
picture. 
\label{parton}}
\end{figure}

\begin{figure}[p]
\centerline{
\epsfig{file=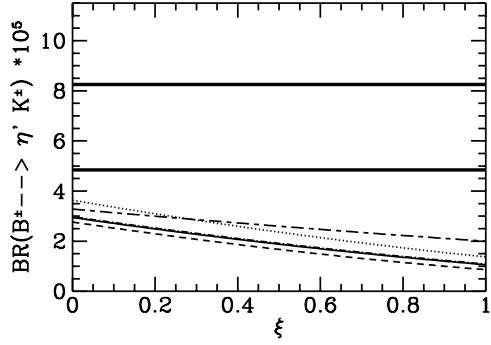,
height=2.0in,angle=0,clip=}
}
\caption[]{
The branching ratio $BR(B^\pm \to \eta^\prime K^\pm)$
plotted against the parameter $\xi$. The lower three curves correspond
to the value $\overline{m_s}( 2.5 ~\mbox{GeV}) = 122 $ MeV and
the three choices of the CKM parameters:
 $\rho = 0.05, \eta = 0.36$ (solid curve);
$\rho = 0.30, \eta = 0.42$ (dashed curve);
$\rho = 0, \eta = 0.22$ (dashed-dotted curve).
The upper two curves correspond to the value $\overline{m_s}( 2.5 
~\mbox{GeV}) = 100 $ MeV, $\rho = 0.05, \eta = 0.36$ and
$f_{\eta^\prime}^c  =-2.3$
MeV from the QCD-anomaly method (dotted curve) and $f_{\eta^\prime}^c
=-5.8$ MeV from \protect\cite{AG97} (long-short dashed curve). The
horizontal thick solid lines represent the present CLEO measurements
(with $\pm 1 \sigma$ errors). 
\label{fig2}}
\end{figure}

\begin{figure}[p]
\centerline{
\epsfig{file=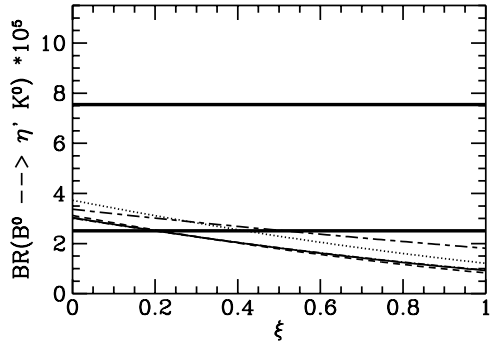,
height=2.0in,angle=0,clip=}
}
\caption[]{
The branching ratio $BR(B^0 \to \eta^\prime K^0)$ plotted
against the parameter $\xi$. 
The legends are the same as in Fig.~3, and the
horizontal thick solid lines represent the present CLEO measurements
(with $\pm 1 \sigma$ errors). 
\label{fig3}}
\end{figure}

\newpage

\section*{Acknowledgments}
The author thanks the organizing committee of this workshop, H.Y. Cheng, 
W.-S. Hou, H.-N. Li and G.-L. Lin for the inivitation. 
He also thanks A. Ali, J. Chay and C. Greub for enjoyable collaboration on 
the subject presented in this talk.  
This work is supported in part 
by the Korea Science and Engineering Foundation (KOSEF)
Contract No. 971-0201-002-2,
the Ministry of Education grant BSRI 98-2418,   
the KOSEF through the SRC program of SNU-CTP, 
and the Distinguished Scholar Exchange Program of Korea Research Foundation.


\section*{References}
\begin{thebibliography}{99}
\bibitem{CLEO1} J. Smith (CLEO Collaboration), talk presented at the
1997 ASPEN winter conference on Particle Physics, Aspen, Colorado,
1997 ; 
S. Anderson et al.~(CLEO Collaboration), CLEO CONF 
97-22a and EPS 97-333 (1997).

 \bibitem{CLEOkpi}
R. Godang et al. (CLEO Collaboration), preprint CLNS 97-1522, CLEO
97-27 (1997).

\bibitem{CLEOomegaphi}
M.S. Alam et al. (CLEO Collaboration), CLEO CONF 97-23 and EPS 97-335
(1997). 

\bibitem{CLEO2}
J. Roy (CLEO Collaboration), invited talk at the Heavy Flavour
Workshop, Rostock, September 1997.

\bibitem{AG97}
A. Ali and C. Greub, Phys. Rev.{\bf D 57}, 2996 (1998).  

\bibitem{AS97}
D. Atwood and A. Soni, Phys.~Lett. {\bf B405} (1997) 150 and 
Phys. Rev. Lett. {\bf 79}, 5206 (1997). 

\bibitem{HZ97}
I. Halperin and A. Zhitnitsky, Phys.~Rev. {\bf D56} (1997) 7247; 
Phys. Rev. Lett. {\bf 80}, 438 (1998). 

\bibitem{SZ97}
E. Shuryak and A.R. Zhitnitsky, 
Phys. Rev. {\bf D 57}, 2001 (1998). 

\bibitem{HT97}
W.S. Hou and B. Tseng,  Phys. Rev. Lett. {\bf 80}, 434 (1998).
 
\bibitem{YC97}
F. Yuan and K.T. Chao, Phys.~Rev. {\bf D56} (1997) 2495.

\bibitem{DHP97}
A. Datta, X.-G. He and S. Pakvasa, 
Phys. Lett. {\bf B419}, 369 (1998). 

\bibitem{CT97}
H.-Y. Cheng and B. Tseng, Phys.Lett.{\bf B415}, 263 (1997). 

\bibitem{KP97}
A. Kagan and A.A. Petrov~, UCHEP-27, UMHEP-443, hep-ph/9707354.

\bibitem{DGR97}
A.S. Dighe, M. Gronau and J. Rosner, Phys.~Rev.~Lett. {\bf 79} (1997)
4333. 

\bibitem{DDO97}
N.G. Deshpande, B. Dutta and S. Oh, 
 Phys. Rev. {\bf D 57}, 5723 (1998). 

\bibitem{Mohd97}
M.R. Ahmady, E. Kou and A. Sugomoto, 
Phys. Rev. {\bf D 58}, 014015 (1998). 

\bibitem{Kim97} 
D. Du, C.S. Kim and Y. Yang, Phys. Lett. {\bf B426}, 133 (1998). 

\bibitem{ACGK} A. Ali, J. Chay, C. Greub and P. Ko, Phys. Lett. {\bf B 424},
161 (1998).

\bibitem{simma}
H. Simma and D. Wyler, Nucl. Phys. {\bf B344}, 283 (1990).

\bibitem{voloshin80}
M. Voloshin and V. Zakharov, Phys. Rev. Lett. {\bf 45} (1980) 688;\\
P. Ball, J.-M.. Fr\`{e}re and M. Tytgat, Phys. Lett. {\bf B365} (1996)
367. 

\bibitem{Leutwyler97}
H. Leutwyler,  Nucl. Phys. Proc. Suppl.{\bf 64} 223 (1998) ;
P. Herrera-Sikoldy, J.I. Latorre, P. Pascual and J. Taron,
Phys. Lett. {\bf B 419}, 326 (1998). 

\bibitem{GK87}
F.J. Gilman and R. Kaufman, Phys.~Rev. {\bf D36} (1987) 2761. 

\bibitem{FK97}
T. Feldmann and P. Kroll, preprint WUB 97-28, hep-ph/9711231.

\bibitem{FKS98}
T. Feldmann, P. Kroll and B. Stech, WU-B-98-2, hep-ph/9802409.
 
\bibitem{AMT98}
F. Araki, M. Musakhanov, H. Toki, Oaksa University Preprint, hep-ph/9803356. 

\bibitem{W83}
L. Wolfenstein, Phys.~Rev.~Lett. {\bf 51} (1983) 1945.

\end{thebibliography}
\end{document}